\documentclass[aps,prb,twocolumn,showpacs,superscriptaddress,10pt]{revtex4}
\usepackage{graphicx}
\usepackage{subfigure}
\usepackage{amsmath}
\usepackage{amsfonts}
\usepackage{amssymb}
\usepackage{amsthm}
\usepackage{times}
\usepackage{soul}
\usepackage[colorlinks,citecolor=blue,linkcolor=blue]{hyperref}

\usepackage{color}

\begin{document}
\title{Triggering waves in nonlinear lattices:\\
Quest for anharmonic phonons and corresponding mean free paths}

\newcommand{\LS}[2]{\textcolor{blue}{#1} \textcolor{red}{\st{#2}}}
\newcommand{\LSS}[1]{\textcolor{magenta}{#1}}


\newcommand{\NUS}{\affiliation{Department of Physics and Centre for Computational Science
and Engineering, National University of Singapore, 117546 Singapore}}
\newcommand{\NGS}{\affiliation{NUS Graduate School for Integrative Sciences and
Engineering, 117456 Singapore}}
\newcommand{\Augsburg}{\affiliation{Institute of Physics, University of Augsburg,
Universit\"atsstr. 1, D-86159 Augsburg, Germany}}
\newcommand{\NIM}{\affiliation{Nanosystems Initiative Munich, Schellingstr, 4, D-80799 M\"unchen,
Germany}}
\newcommand{\Tongji}{\affiliation{Center for Phononics and Thermal Energy Science,
School of Physics Science and Engineering, Tongji University, 200092 Shanghai, China}}
\newcommand{\USA}{\affiliation{Theoretical Division, Los Alamos National Laboratory, Los
Alamos, 87545 New Mexico, USA}}
\newcommand{\Fudan}{\affiliation{State Key Laboratory of Surface
Physics and Department of Physics, Fudan University, 200433 Shanghai,
China}}
\newcommand{\Graphene}{\affiliation{Graphene Research Centre, Faculty of Science, National
University of Singapore, 117542 Singapore}}

\author{Sha Liu}
\email{phylius@nus.edu.sg}
\NUS
\NGS
\author{Junjie Liu}
\Fudan
\NUS
\author{Peter H\"anggi}
\email{hanggi@physik.uni-augsburg.de}
\Augsburg
\NIM
\NUS
\Tongji
\author{Changqin Wu}
\Fudan
\author{Baowen Li}
\email{phylibw@nus.edu.sg}
\NUS
\NGS
\Tongji
\Graphene

\date{17 Nov 2014}

\newcommand{\vdot}{}
\newcommand{\vect}{}
\newcommand{\vct}{\boldsymbol}
\newcommand{\vp}{\vect{p}}
\newcommand{\vx}{\vect{x}}
\newcommand{\vv}{\vect{v}}
\newcommand{\vF}{\vect{F}}

\newcommand{\EPT}{EPT}
\newcommand{\fpu}{FPU}
\newcommand{\fpuab}{FPU-$\alpha\beta$}
\newcommand{\fpub}{FPU-$\beta$}

\newcommand{\lpt}[2]{\frac{\partial #1}{\partial #2}}
\newcommand{\ilpt}[2]{\partial_{#2}{#1}}
\newcommand{\lppt}[3]{\frac{\partial^{#1} #2}{\partial #3^{#1}}}
\newcommand{\OL}[1]{\mathcal{L}^{{#1}}}
\newcommand{\bkb}[1]{\left(#1\right)}
\newcommand{\bks}[1]{\left[#1\right]}
\newcommand{\fracc}[2]{{#1}/{#2}}
\newcommand{\RP}{{R-ph}}
\newcommand{\AP}{{a-ph}}

\newcommand{\figref}[1]{Fig.~\ref{#1}}
\newcommand{\expk}{\alpha}
\newcommand{\expx}{\beta}
\newcommand{\Sec}[1]{{\bf #1}}
\newcommand{\levy}{L\'evy}
\renewcommand{\Omega}{\omega}
\renewcommand{\Delta}{}

\newcommand{\mean}[1]{\left\langle#1\right\rangle}
\newcommand{\dv}[1]{\delta\!\mean{#1}_{neq}}   
\newcommand{\non}[1]{\mean{#1}_{neq}}  
\newcommand{\eqi}[1]{\mean{#1}_{eq}}        

\newcommand{\NN}{\mathcal{N}}
\newcommand{\define}{\equiv}
\newcommand{\msdp}{\langle x_p^2(t)\rangle}
\newcommand{\msd}{\langle \Delta x^2(t)\rangle_E}
\newcommand{\Corr}{\mathcal{C}}
\newcommand{\bt}{\beta_T}
\newcommand{\textemph}[1]{\textcolor{blue}{\bf #1}}

\newcommand{\mfp}{{\ell}}
\newcommand{\hf}{\frac{1}{2}}
\newcommand{\RE}{\mathrm{Re}}
\newcommand{\saba}{$c$SABA$_2$}
\newcommand{\meann}[2]{\left\langle#2\right\rangle_{#1}}
\newcommand{\dd}{d}
\newcommand{\intinf}{\int_0^\infty}

\begin{abstract}
Guided by a stylized experiment we develop a self-consistent anharmonic phonon concept for
nonlinear lattices which allows for explicit ``visualization.'' The idea uses a small external driving force which
excites the front particles in a nonlinear lattice slab and subsequently one monitors  the excited wave evolution using molecular dynamics simulations. This allows for a simultaneous, direct determination of the existence of the phonon
mean free path with its corresponding anharmonic phonon wavenumber as a function of temperature.
The concept for the mean free path is very distinct from known prior approaches: the latter  evaluate
the mean free path {only} indirectly, via using both,   a scale 
for the phonon relaxation time and yet another one for the phonon velocity. Notably, the concept here is
neither limited to small lattice nonlinearities nor to small frequencies. The scheme is tested for three  strongly  nonlinear
lattices of timely current interest  which either exhibit normal or anomalous heat transport.
\end{abstract}
\maketitle

\section{Introduction}
In solid phases, phonons are collective, elementary
vibrations in {harmonic} lattices and as such play a prominent  role for
physical transport phenomena aplenty
 \cite{Born.54.NULL,Peierls.29.AP,Peierls.55.NULL,Ziman.72.NULL}, of which the transport
of heat is a most prominent one.  In harmonic lattices these phonons constitute nondecaying, stable propagating waves
obeying a dispersion relation for angular frequency $\omega$ and  corresponding wavenumber
$k$. This in turn implies that these phonons possess strictly infinite mean free paths (MFPs).
Consequently, heat transport in harmonic lattices is ballistic  \cite{Rieder.67.JMP,Segal.03.JCP}
and thus no temperature gradient  can be sustained (breakdown of Fourier's law).

Generally, however, everyday solid materials are far from being perfect harmonic
lattices. Therefore, the phonon concept bears no firm basis {{\it away}} from its underlying (effective) harmonic  approximation.
As pointed out by Peierls long ago \cite{Peierls.29.AP,Peierls.55.NULL}, such anharmonicity is
essential for Umklapp scattering --- an indispensable process for a finite,
size-independent thermal conductivity $\kappa$ in three dimensional (3D)
materials. Phenomenologically \cite{Peierls.55.NULL,Ziman.72.NULL}, the  thermal conductivity is
approximated in terms of a wave-number-dependent  phonon MFP $l_k$; i.e. $\kappa =
(1/3)\sum_k C_{k} v_k l_k$. Here, $C_k$ is the specific heat of the phonon mode
and $v_k$ its phonon group velocity.

\begin{figure}[t]
\includegraphics[width=0.75\columnwidth]{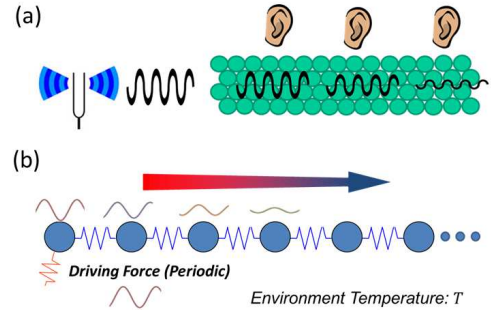}
\caption{(color online) Hunting for phonons and MFPs in nonlinear lattices: (a) An illustration of
the
``tuning fork experiment''; (b) a
schematic sketch of the driving force method in a nonlinear lattice.}
\label{fig:model}
\end{figure}

Principally, we encounter the dilemma  that the rigorous existence of a phonon excitation in a nonlinear lattice is
self-contradictory to the very existence of a finite MFP. Particularly, this concept of a nonlinear or anharmonic  phonon
may cause considerable unease when dealing with strong nonlinear
interaction forces and/or high temperatures where thermal
excitations  no longer predominantly dwell the harmonic well regions of  corresponding interaction
potentials.  Moreover,
the observed  breakdown of Fourier's law  with
superdiffusive heat transport in systems of low dimensions \cite{Lepri.03.PR,Dhar.08.AP,Liu.12.EPJB}
with a thermal conductivity diverging with increasing length of the sample necessitates that some MFPs must diverge.

One method  addressing the issue is the so termed
renormalized phonon picture
\cite{Alabiso.95.JSP,Alabiso.01.JPA,Gershgorin.05.PRL, Gershgorin.07.PRE,
Li.06.EL,Li.10.PRL}. In essence, this approach uses an effective harmonic approximation  of the
nonlinear interaction forces via a temperature-renormalized phonon dispersion. {Such renormalized phonon theory, however, neither  determines the  phonon MFP nor a phonon relaxation time (or, likewise, a phonon lifetime). The problem of finite MFPs and corresponding relaxation times thus remains open.} One possibility addressing the missing link consists in combining  approximate
Boltzmann transport theory for heat transport  with a  single mode relaxation time approximation
\cite{Callaway.59.PR,Holland.63.PR}. In fact, while this  phenomenological scheme is commonly adopted nowadays,
its regime of validity has never been justified from first principles \cite{Herring.54.PR,Pereverzev.03.PRE}.

{In summary,} the present state of the art is that the physical value for this thermal MFP $\mfp$
is evaluated {{\it indirectly} only:} Its evaluation
involves both a characteristic relaxation time or lifetime scale $\tau$ and as well  a characteristic scale for the phonon
speed $v$, yielding $\mfp=v\tau$. Thus, this  MFP  is not uniquely given in the sense that various time scales
come to mind; namely, the so termed phonon lifetime (as obtained from Lorentzian fit) in prior  phonon quasiparticle
studies in the reciprocal lattice space \cite{Henry.08.JCTN,Turney.09.PRB,Thomas.10.PRB,Sun.10.PRB,Zhang.14.PRL,Maradudin.62.PR,Collins.78.PRB,Chaikin.book.2000,Pang.13.PRL},
{a phonon collision time or transport relaxation time (notably being not equivalent with the phonon lifetime) as  obtained from a Peierls--Boltzmann approach are just but a few \cite{Ward.10.PRB,Sun.10.PRBa}. Likewise, the {\it a priori} choice for the speed scale taken as the group velocity is also empirical.
-- Consequently, a  direct, molecular dynamics (MD) based visualization rendering this much sought-after, physically descriptive
and useful phonon MFP is  highly desirable.}

{Here,} we put forward an experiment-inspired theoretical scheme for
propagating anharmonic phonons (\AP{}s) when  ubiquitous  nonlinear interaction
forces are ruling the lattice dynamics.  Our main objective is to significantly advance an
\AP{} concept that simultaneously solves the following challenges:
(i) the concept can be evaluated by MD simulations and additionally allows for the  visualization of the physical
existence of the \AP{}'s together with their MFPs, (ii) the concept is manifest  nonperturbative
in the strength of lattice nonlinearities, and  additionally, (iii) the concept is
neither restricted to low temperature nor to low frequencies.

\section{ Triggering anharmonic phonons.}

To elucidate whether a  phonon picture still holds in
strongly nonlinear solids
we use a ``tuning fork experiment'' as sketched with \figref{fig:model}(a).
Here, a tuning fork  operating at small driving strength and at a  fixed frequency $\Omega$
is placed in front of a crystalline, nonlinear lattice slab held at  a  temperature $T$.
This  driving source will generate  sound that propagates
along the slab. For a phonon picture to hold up, it is then required
that the propagating disturbance physically
causes a collective,  attenuated plane wave-like response.
For the MFP to exist, the spatially dependent wave amplitude  preferably is required to exhibit
an exponential decay with a single scale vs increasing spatial spread.
If so, this renders  the sought-after MFP for \AP{}'s in a nonlinear
lattice. The \AP{} may still  hold up, however, even if the attenuation of the propagating
wave occurs non-exponentially, i.e., when exhibiting   multiple
spatial scales, see below.

The driving source triggering the thermal phonons must be set sufficiently
small so that no nonlinear, non-phonon-like excitations become excited.
This implies that the triggered response of the tuning fork occurs solely within its linear regime, i.e.,  the output
signal occurs at the same (driving) frequency only \cite{Hanggi.82.PR}.

To realize this stylized experiment, we start with the lattice being held at thermal
equilibrium. We next apply an time-dependent external weak force $f_d(t)=f_1\cos\Omega t$, see  in
\figref{fig:model}(b), to the
first particle and measure the resulting long time response occurring at all the remaining particles.
For the \AP{} concept to make sense this collective response must
assume the form of a propagating plane
wave, i.e., the thermally averaged velocity $v_n(t)$ of the $n$-th particle is required to read for $n=1,2,\cdots$:
\begin{equation}\label{eq:wave}
\mean{v_n(t)}_f = |A_n| \cos(\omega t+\phi_n)=\RE (|A_n| e^{i(\phi_n+\omega
t)}),
\end{equation}
with the phase obeying
\begin{equation}
\phi_n=-kn+\phi_0 .
\label{phaserelation}
\end{equation}
In the expression, $\mean{\cdot}_f$ denotes the statistical average under the influence of the driving
force. This so parameterized excited motion defines an effective  phonon  with a frequency $\Omega$
that precisely matches the input driving frequency $\Omega$.
The coefficient, $k=-d\phi_n/dn$, plays the role of the wavenumber $k$.  With the
amplitude $|A_n|$ assumed to decay exponentially as
\begin{equation}
 |A_n| \propto e^{-n/\mfp},
\end{equation}
its  decay length $\mfp$  provides  the searched MFP for
this  \AP{}.

For the sake of simplicity,
we first formulate the concept for one-dimensional (1D) lattices.
 The scheme can readily be generalized to higher
dimensions and complex materials by applying forces to atoms  lying on a
chosen lattice plane that trigger either longitudinal or transverse waves which propagate
perpendicular to the plane, which will be discussed in Sec. \ref{sec:high_D}.

The 1D lattice
Hamiltonian assumes the general dimensionless  form \cite{Unit.00.NULL}:
\begin{equation}
	H_0=\sum_{n=1}^N \bks{\frac{p_n^2}{2}+ V(x_{n+1}-x_{n})+U(x_n)}\;,
\end{equation}
where $p_n$ and $x_n$ denote the momentum and displacement from
the equilibrium position for the $n$-th particle (with unit mass),
respectively, $V(x_{n+1}-x_{n})$ is the inter-particle potential and $U(x_n)$
denotes a possibly present on-site potential.
Following the common approach we employ  the periodic
boundary conditions.
The role of finite temperature $T$ enters by using  a canonical ensemble
with the unperturbed distribution reading,
$\rho_{eq}= Z^{-1} \exp {[-\beta_{T}  H_{0}]}$, where $\beta_T=1/k_B T$ is the inverse temperature
and $Z$ the canonical partition function.

Following the spirit of the ``tuning fork experiment'', we then apply a weak single--frequency signal $f_d(t)$ to the first particle of the lattice. Therefore, the total Hamiltonian of the system reads
\begin{equation}
H_{{tot}}= H_0 + H_{{ext}}= H_0 - f_d(t) x_1.
\end{equation}

 We can then calculate the thermally averaged velocity at each site $n$ using
 canonical linear response theory \cite{Hanggi.82.PR,Kubo.91.NULL}, yielding:
\begin{equation}\label{eq3:v}
\begin{split}
\mean{v_n(t)}_f&=\beta_T \int_{-\infty}^t ds
\mean{v_n(t-s)v_1(0)} f_d(s), \\
&= \beta_T \int_{0}^{\infty} d\tau
\mean{v_n(\tau)v_1(0)} f_d(t-\tau)
\end{split}
\end{equation}
where $\mean{\cdot}$ denotes the
canonical ensemble average.

For $f_d(t)=f_1\cos\Omega t$,
the excited motion can equivalently be cast into the form of Eq. \eqref{eq:wave}, which  involves
the  Fourier transformed  susceptibility ,
i.e.,
\begin{equation}
\mean{v_n(t)}_f= f_1 \RE[{\chi_n(\omega)e^{i\Omega t}}],
\label{eq:R}
\end{equation}
where the susceptibility $\chi(\Omega)$ reads
\begin{equation}
\label{eq:chi}
\chi_{n}(\Omega)\!=\! \beta_T\! \int_0^\infty\! d{\tau}
\mean{v_n(\tau)v_1(0)}e^{-i\Omega \tau}\equiv |\chi_n|e^{i
\phi_n}.
\end{equation}

This appealing result  allows one to assign the
existence of an \AP{} and its corresponding MFP: (i) an \AP{} exists with a wavenumber
$k(\Omega)$ if and only if the linear relationship in \eqref{phaserelation} is fulfilled
and (ii) possesses a unique  MFP $\mfp(\Omega)$ when
 $|\chi_n(\Omega)|\propto \exp [-n/\mfp(\omega)]$. Importantly,
 the phonon response amplitude $|\chi_n(\omega)|$ and its phase $\phi_n(\omega)$ now both attain a dependence
 on temperature $T$. The wavevector is given by $k(\omega,T)=-d\phi_{n}(\omega,T)/dn$. This very form considerably simplifies the
numerical efforts as compared to directly studying the excited waves via the MD method, i.e., one
finds the whole frequency-resolved phonon properties at once.

As an expectation, this so introduced \AP{} concept should be naturally consistent with the normal phonon in harmonic lattices. Such a fact can be readily tested analytically without invoking the linear response theory, which we demonstrate below. Following this, we give a brief extension of our method to the more general three dimensional case and then move on to the application of our method.

\subsection{Harmonic Lattices}
For a harmonic lattice with potentials $V(x)= \frac{1}{2} x^2$ and $U(x)=0$, we expect to observe
from our method that the MFPs are infinite with the wavenumber $k$ satisfying
\begin{equation}
\label{eq:harmonic_disp}
k(\omega)= 2 \arcsin \frac{\omega}{2}.
\end{equation}

To see this, we apply periodic boundary conditions to the lattice so that $x_{0/1}= x_{N/N+1}$ and $p_{0/1}= p_{N/N+1}$.
With a periodic driving force $f_d= f_1\cos \Omega t$ switched in the infinite past and applied to the first particle in a 1D-chain, the
equations of motion (EOMs) can be put into a compact matrix form, reading
\begin{equation}
\label{eq3:eom_ha}
	\ddot{\vct{x}}=- \mathbf{\Phi}\vct{x}+ \vct{F}(t).
\end{equation}
Here, $\mathbf{\Phi}$ is the force matrix with elements
$\mathbf{\Phi}_{i,j}=2\delta_{i,j}- \delta_{i,j-1}-\delta_{i,j+1}$ in terms of the Kronecker delta
function $\delta_{i,j}$, and $\vct{F}=\vct{f}
\cos(\Omega
t)=(f_1,0,\cdots,0)^T\cos(\Omega t)$.
Its  solution is additive due to  $\vct{F}(t)$ entering a linear equation of motion. Thus, the excitations of
$\vct{F}(t)$ can be obtained by Fourier transformation, reading
\begin{equation}
\label{eq3:sol1}
	\mean{\vct{x}(t)}_f=\int_{-\infty}^{\infty}\mathbf{G}(\omega') \tilde{\vct{F}} (\omega') e^{i\omega' t}d\omega',
\end{equation}
where $\tilde{\;}$ denotes a Fourier transform and $\mathbf{G}$ is the phonon Green's function
\begin{equation}
\label{eq3:green}
	\mathbf{G}(\omega')= \bkb{\mathbf{\Phi}-\omega'^2}^{-1}.
\end{equation}

\newcommand{\expo}[1]{e^{#1 i\omega t}}
For a  driving $f_d= f_1 \cos{\omega t}$, it can then be calculated that the resulting excited motion reads
\begin{equation}
\mean{x_n(t)}_f=f_1 \RE \left[\mathbf{G}_{n,1}(\omega)e^{i\omega t}\right],
\end{equation}
and consequently \begin{equation}
\label{eq:excited}
	\mean{v_n(t)}_f= f_1\RE\left[i\omega\mathbf{G}_{n,1}(\omega) e^{i\omega t}\right],
\end{equation}
which has the same form as \eqref{eq:R} with $\chi_n(\omega)= i\omega \mathbf{G}_{n,1}$.

Due to the cyclic structure of the matrix $\mathbf{\Phi}-\omega^2$, its inverse $\mathbf{G}$ can be analytically obtained. Its first column reads
\begin{equation}
\label{eq3:g}
    \mathbf{G}_{n,1}=-\frac{\cos\bkb{\frac{N}{2}-n+1}z}{2\ \sin(Nz/2)\ \sin z},
\end{equation}
The second column is obtained by cyclically shifting the first column by one element. The third column
is obtained by cyclically shifting the first column by two elements, and so on.
In the formula,  $e^{\pm i z}$ are the two roots of the quadratic equation $
	-1+ (2-\Omega^2) x- x^2=0
$. Therefore, $z$ satisfies
\begin{equation}
\label{eq3:d1}
	\cos z = 1-\frac{\Omega^2}{2}\;,
\end{equation}
which can be verified by substitution.

For
$0<\omega<2$, $z$ is a real number. According to \eqref{eq:excited} and \eqref{eq3:g}, the velocity of excited wave
varies with $n$ as
\begin{equation}
	\label{eq:not_eigen}
	\mean{v_n(t)}_f\sim \cos\left[\big(\frac{N}{2}-n+1\big) z \right] \sin\omega t\;.
\end{equation}
It represents a standing wave formed by two
plane waves with the same wavenumber $k=z$ satisfying
\begin{equation}
	\omega=2\sin\frac{k}{2},
\end{equation}
which just yields the intrinsic dispersion relation for the harmonic lattice. The plane wave nature also implies the excited waves have infinite MFPs.

Note that the derivation above is exact and beyond a linear response. However, it can be shown that applying \eqref{eq:R} and \eqref{eq:chi} to the harmonic lattice will give exactly the same result in \eqref{eq:excited} and \eqref{eq3:g}, due to the fact that for harmonic lattices all higher order response terms are actually zero and only the linear response term is present.

\subsection{Generalization to higher dimensions\label{sec:high_D}}

In order to generalize our method to higher dimensional cases, we must apply a periodic driving force
at the same frequency to each particles in a plane. Taking the three dimensional (3D) case as example, if we want to study the wave propagation
along the direction ${\bf{k}}=h \mathbf{b_1}+k\mathbf{b_2}+ l \mathbf{b_3}$ where $\bf{b_1}$,
$\bf{b_2}$, and
$\bf{b_3}$ are the primitive vectors in the reciprocal lattice, we  first choose a plane
with Miller indices $(hkl)$, denoted as $\alpha$, being orthogonal to the direction
${\bf{k}}$. Then, we apply forces to all particles in this plane. To trigger longitudinal
waves, we apply out-of-plane forces that are perpendicular to the plane. To trigger transverse
waves, we apply in-plane forces.

With such an setup, \eqref{eq3:v} can be easily generalized to read
\begin{equation}
\label{eq3:v2}
	\meann{f}{v_n^d(t)}= \bt\sum_{i\in\alpha} \intinf \dd{\tau}\  \mean{v_n^d(\tau)v_i^d(0)}
f_i^d(t-\tau),
\end{equation}
where $i\in \alpha$ means particle $i$ is in plane $\alpha$ and $d= \perp$ or $\parallel$ which specifies whether the direction of the force and velocity is
out-of-plane ($ \perp$) or in-plane ($\parallel$). We now identically set $f_i^d(t)= f_\alpha^d(t)/N_\alpha$ ($N_\alpha$ is
the number of particles
in that plane) for all $i\in\alpha$; then \eqref{eq3:v2} can be simplified to read
\begin{equation}
\label{eq3:v3}
	\meann{f}{v_n^d(t)}= \bt \intinf \dd{\tau}\  \mean{v_n^d(\tau)v_\alpha^d(0)} \
f_\alpha^d(t-\tau),
\end{equation}
where $v_\alpha^d=(1/N_\alpha)\sum_{i\in\alpha}v_i^d$ denotes the average velocity for all particles
in the plane $\alpha$.

We can further take an average for all particles in the same lattice plane,
denoted as $\beta$, which has the same Miller index (so that it is parallel to $\alpha$) but
contains the $n$-th particle. Therefore, we obtain for the average velocity for that very plane the result
	\begin{equation}
	 \label{eq3:v4}
	\meann{f}{v_\beta^d(t)}= \bt \intinf \dd{\tau}\  \mean{v_\beta^d(\tau)v_\alpha^d(0)}
f_\alpha^d(t-\tau).
\end{equation}
From this result we can infer whether an anharmonic phonon  with a wavevector pointed towards the
same direction of ${\bf{k}}$ exists or not,  by following the same reasoning used  for one dimensional lattices.

Note that the derivation is independent of the form of the Hamiltonian $H_0$. So the method is equally
applicable to study wave transport in inhomogeneous lattices, such as junctions formed by different
materials.

\section{Numerical details}
We will apply our concept to three archetype 1D nonlinear lattices  of varying complexity.
Before we move on to study these models in detail, we first describe the numerical details used to detect
 the anharmonic phonons.

We use throughout a fourth order symplectic \saba{} algorithm
 to integrate the Hamiltonian equations of  motion \cite{Laskar.01.CMDA}. The time
step has always been chosen as $h=0.02$ and the length has been set at  $N=2048$, using periodic boundary conditions, for all models studied. At the
beginning of each  simulation, a total time $t=2\times 10^6$ is used to thermally equilibrate the
system.
After that, the time-homogeneous equilibrium velocity correlation $\mean{v_n(t)v_1(0)}$ is calculated by using the time average  which replaces,
using ergodicity for the nonlinear lattice,  the corresponding ensemble average. An average over
$3.2\times 10^{9}$ steps is used. The correlation
$\mean{v_n(t)v_1(0)}$ is calculated for each $n=1,2,\cdots,N$ and
$t=0,h,2h,\cdots,t_{\mathrm{max}}$. Afterwards, ${\chi}_n(\Omega)$ is
obtained by taking a Fourier transform according to Eq. (7).

 \begin{figure}[tb!]
 \centerline{
 \includegraphics[width=0.4\textwidth]{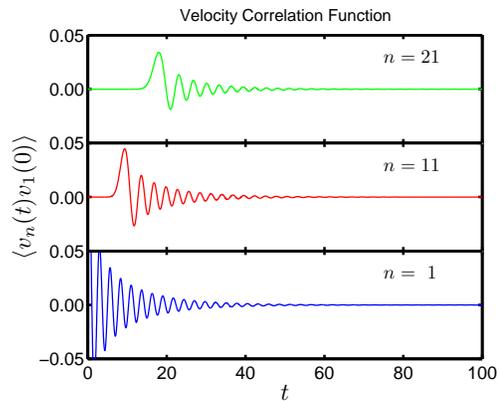}
 }
  \caption[The velocity auto-correlation of the \fpub{} model]{(Color online) The velocity
 auto-correlation $\mean{v_n(t)v_1(0)}$ of the \fpub{} model for $n=1,11,21$. The simulation is carried out on
 an \fpub{} lattice with length N=2048 at temperature $T=0.2$.}
  \label{fig3:fpu.vv}
 \end{figure}

The upper time limit $t_{\mathrm{max}}$ is properly chosen such that the excited waves along the
periodic ring  of size $N$ do not overlap with each other for $t\in (0, t_{\mathrm{max}})$. Namely,
$t_{\mathrm{max}}< N/2v_s$ where $v_s$ is the largest group velocity of the phonons studied, i.e.,
the corresponding sound velocity. On the other hand, this $t_{\mathrm{max}}$ determines the
frequency resolution. The larger $t_{\mathrm{max}}$ is, the smaller is the frequency resolution.
Specifically, $t_{\mathrm{max}}=655.36$ has been used for all three models.

A few samples of the velocity correlation $\mean{v_n(t)v_1(0)}$ are depicted in Fig. \ref{fig3:fpu.vv} for the Fermi-Pasta-Ulam (FPU)-$\beta$
lattice at a temperature $T=0.2$.  $\chi_n(\omega)$ is then calculated via the Fourier transformation
according to Eq. (7).  For the other models, we also observe similar oscillation behavior for the
velocity-velocity correlation function.

\section{ Detecting anharmonic phonons.}
In this section, we apply our method to study three archetype 1D nonlinear lattices.
Of timely interest in the context  of anomalous {\it vs.} normal heat conduction
are the FPU-$\beta$  lattice, the \fpuab{} lattice  and the
$\phi^4$ lattice. Our numerical simulations shall cover
extended regimes of temperature $T$ and frequencies $\Omega$.

\subsection{\fpub{} lattice.}

\begin{figure*}[tb!]
 \centering
 \subfigure[\ $\chi_n(\omega= 1.505)$]{
  \includegraphics[width=0.32\textwidth]{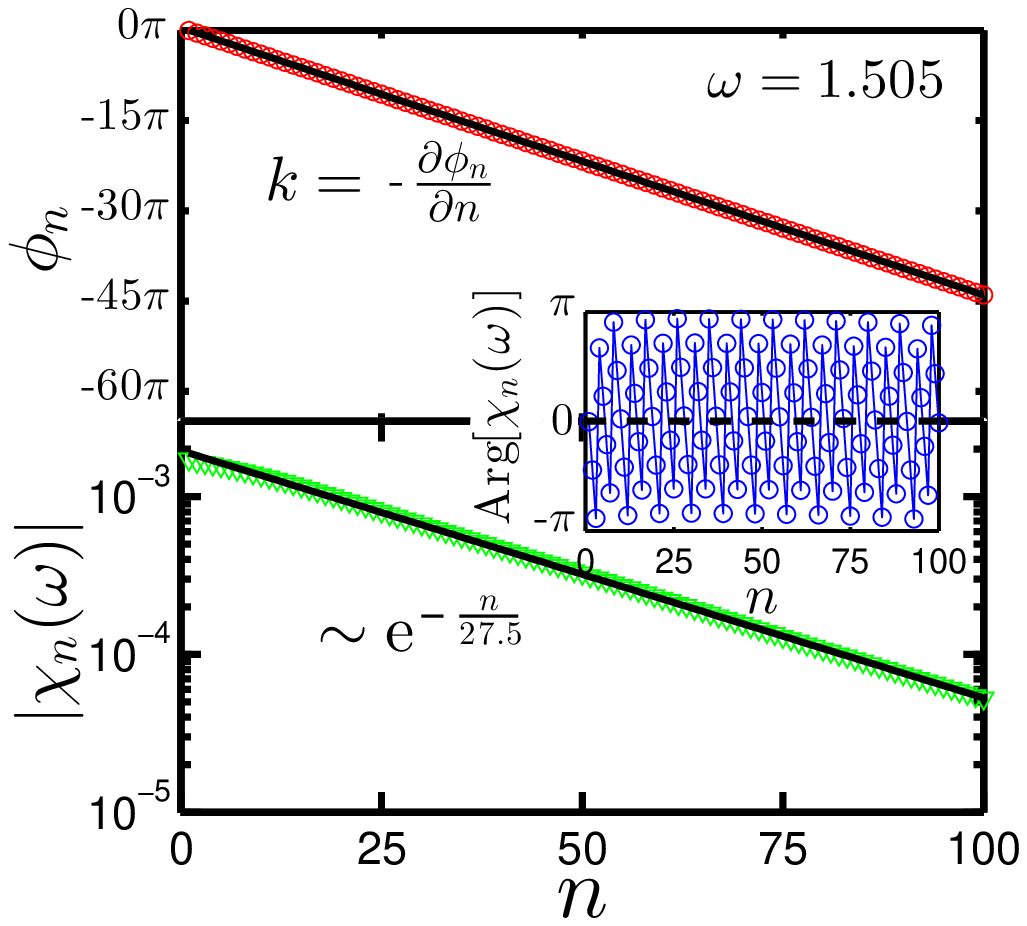}}
   \subfigure[\ The phase $\phi_n$]{
  \includegraphics[width=0.32\textwidth]{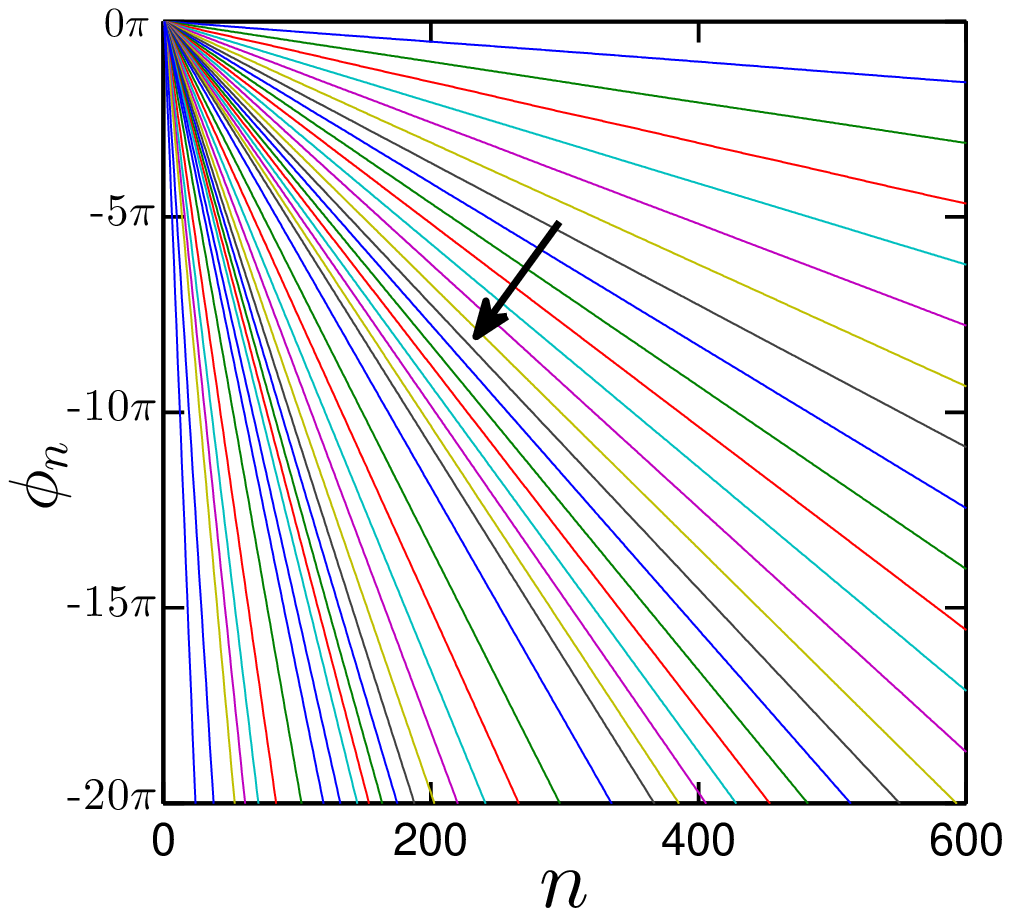}}
     \subfigure[\ the amplitude $|\chi_n|$]{
  \includegraphics[width=0.32\textwidth]{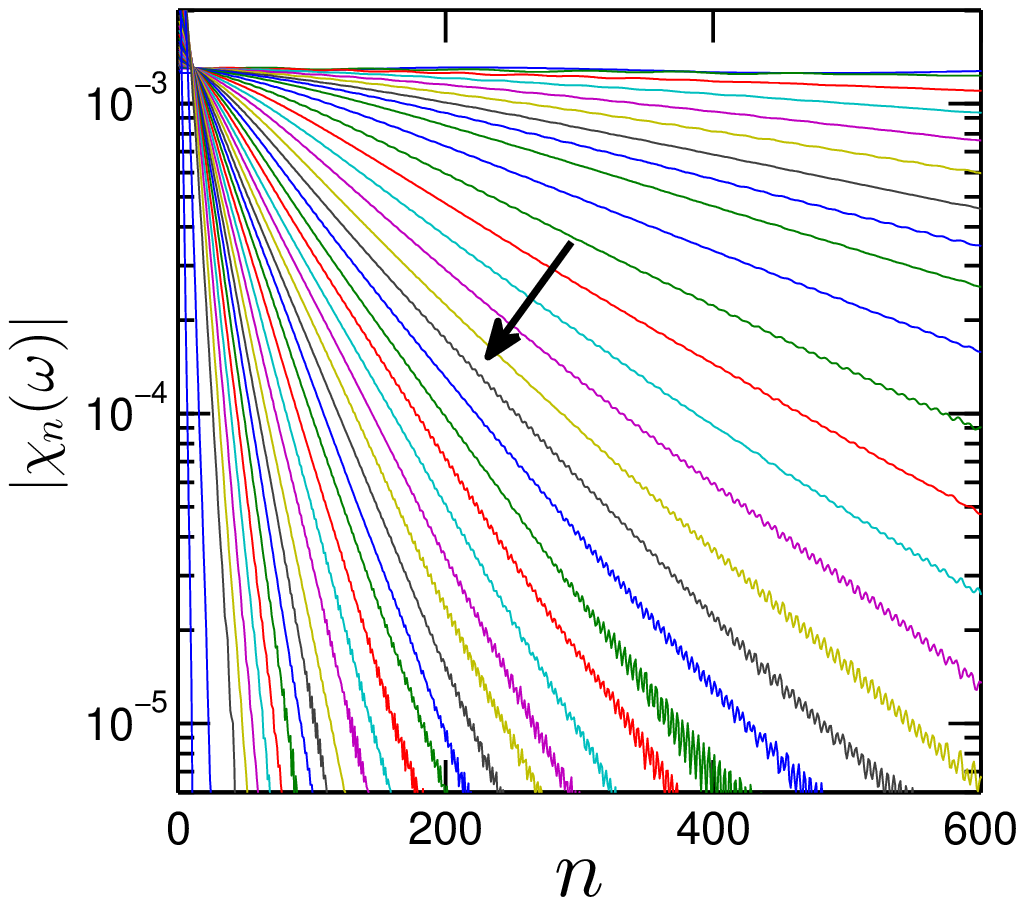}}
\caption{(Color online) The response function for an
FPU-$\beta$ chain with length $N=2048$ at temperature $T=0.2$. (a) A detailed example for $\chi_n(\omega)$ at frequency $\omega= 1.505$ as a function of $n$. The upper panel shows the phase $\phi_n$ and the lower panel shows the amplitude $|\chi_n|$. The inset shows the principal values of the arguments
$\mathrm{Arg}[\chi_n]\in[-\pi,\pi)$. The principal
value jumps discontinuously by $2\pi$ when $-\pi$ is reached. To obtain a continuous varying phase
$\phi_n$, as depicted in the upper panel, we shift the arguments by 2$\pi$ after each such jump.
(b), (c): A comprehensive view of the phase and the amplitude, respectively, for different frequencies $\omega\in (0.0096, 2.675)$.
The driving frequency $\Omega$ increases along the arrow.
}
\label{fig:R_all}
\end{figure*}

\begin{figure}[b!]
\includegraphics[width=0.9\columnwidth]{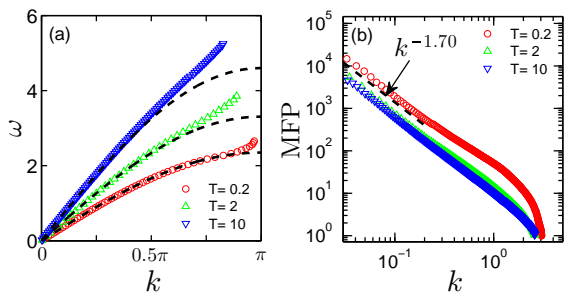}
\caption{(Color online) (a) The dispersion relation for the FPU-$\beta$ model.
The dashed curve are obtained from the
renormalized phonon theory using Eq. \eqref{eq:EPT}. (b)
Corresponding phonon MFPs. The dashed lines are illustrations for a power-law
behavior $\mfp\sim{}k^{-1.70}$.
}
\label{fig:fpu_dm}
\end{figure}

The prevalently studied nonlinear 1D lattice dynamics in the  literature is the \fpub{} dynamics with
$V(\Delta x)= \frac{1}{2}\Delta
x^2+\frac{1}{4}\Delta x^4$ and $U(x)=0$  \cite{Ford.92.PR,Berman.05.C}. Its lattice dynamics has
been demonstrated  to exhibit superdiffusive heat transport
\cite{Lepri.97.PRL,Lepri.03.PR,Dhar.08.AP,Liu.12.EPJB}.

Applying the \AP{} concept,  our findings are depicted in  \figref{fig:R_all} for  $\chi_n(\omega)$ {\it vs.} lattice sites $n$,
for different driving frequencies $\Omega\in (0.0096,2.675)$ and a dimensionless temperature $T=0.2$
\cite{Unit.00.NULL}.
Beyond  $\Omega=2.675$ the response decays very fast, yielding also a very short
phonon MFP. This  limits the evaluation of the corresponding wavenumber $k$---
practically, it cannot be extracted with good confidence near $k\approx\pi$.

The results in \figref{fig:R_all} provide twofold relevant information:

(i) Firstly, for all frequencies depicted, the phase $\phi_n$   perfectly decreases linearly
with $n$.  This  corroborates the existence of  \AP{}'s with a corresponding wavenumber $k=-d\phi_n/dn$.
We evaluate  $k(\Omega)$ for different driving frequencies $\Omega$, as depicted in
\figref{fig:fpu_dm}(a), and compare our results  with
predictions taken from  renormalized phonon theory (dashed lines) \cite{Li.06.EL,Li.10.PRL},
which predicts
\begin{equation}
\label{eq:EPT}
	\omega=2\alpha(T) \sin\frac{k}{2}\;.
\end{equation}
Here, $\alpha(T)$ denotes the temperature-dependent renormalization factor that
quantifies the strength of  nonlinearity. The temperature dependent sound speed  $v_s$ emerges as
$v_s = \frac{d\omega}{dk}|_{k=0}=\alpha(T)$. For the \fpub{} lattice
$\alpha(T)=\big({1+\frac{\int
x^4 e^{-(x^2/2+x^4/4)/T}dx}{\int x^2 e^{-(x^2/2+x^4/4)/T}dx}}\big)^{1/2}$ \cite{Li.06.EL,Li.10.PRL}; it
increases with  temperature, starting out at $1$. As seen in  \figref{fig:fpu_dm}(a), excellent agreement is obtained for
low frequency \AP{}'s. The differences between the \AP{} concept and renormalized phonon
theory occur at large frequencies, with deviations slightly increasing  with increasing temperature.
This corroborates with the fact that effective phonon theory self-consistently applies
to weak anharmonic forces and long wavelength phonons only.

(ii) Secondly, within the depicted frequency regime the amplitude $|\chi_n|$,
\figref{fig:R_all}(c),  perfectly
decays exponentially with increasing $n$. Therefore, a sensible MFP $\mfp$ is obtained
for each \AP{}, cf. in \figref{fig:fpu_dm}(b), where we depict the
MFPs at three different ambient temperatures. Moreover,  the MFPs
diverge with the decreasing  wavenumber $k$. This is a salient feature known for momentum
conserving 1D systems \cite{Lepri.98.EL}.


Interestingly, a power-law divergence
$\mfp(k) \sim k^{-\mu}$ with $\mu\approx 1.70$ is observed for small
$k$ for various temperatures. The numerical result closely matches the prediction of
Peierls--Boltzmann theory at  weak anharmonic nonlinearity, rendering  $\mu=5/3$
\cite{Pereverzev.03.PRE,Nickel.07.JPA,Lukkarinen.08.CPAM}.
A divergent exponent $\mu>1$ causes an  anomalous
divergent heat conductivity $\kappa\sim N^\beta$ with $\beta =
1-1/\mu$ \cite{Pereverzev.03.PRE,Dhar.08.AP}. Therefore, we numerically
find
$\beta\approx 0.411$, which is  close to
results in \cite{Lepri.97.PRL,Wang.11.EL}.

We stress that with our concept of the \AP{} the  existence of
MFPs   (or its corresponding transport relaxation time $\tau_k$) in
the \fpub{} lattice is here not postulated {\it a
priori} \cite{Pereverzev.03.PRE,Nickel.07.JPA,Lukkarinen.08.CPAM} but is  confirmed independently via  MD
simulations in configuration space.

\subsection{\fpuab{} lattice.}

\begin{figure*}[tb]
 \centering
 \subfigure[\ $\chi_n(\omega= 1.505)$]{
  \includegraphics[width=0.32\textwidth]{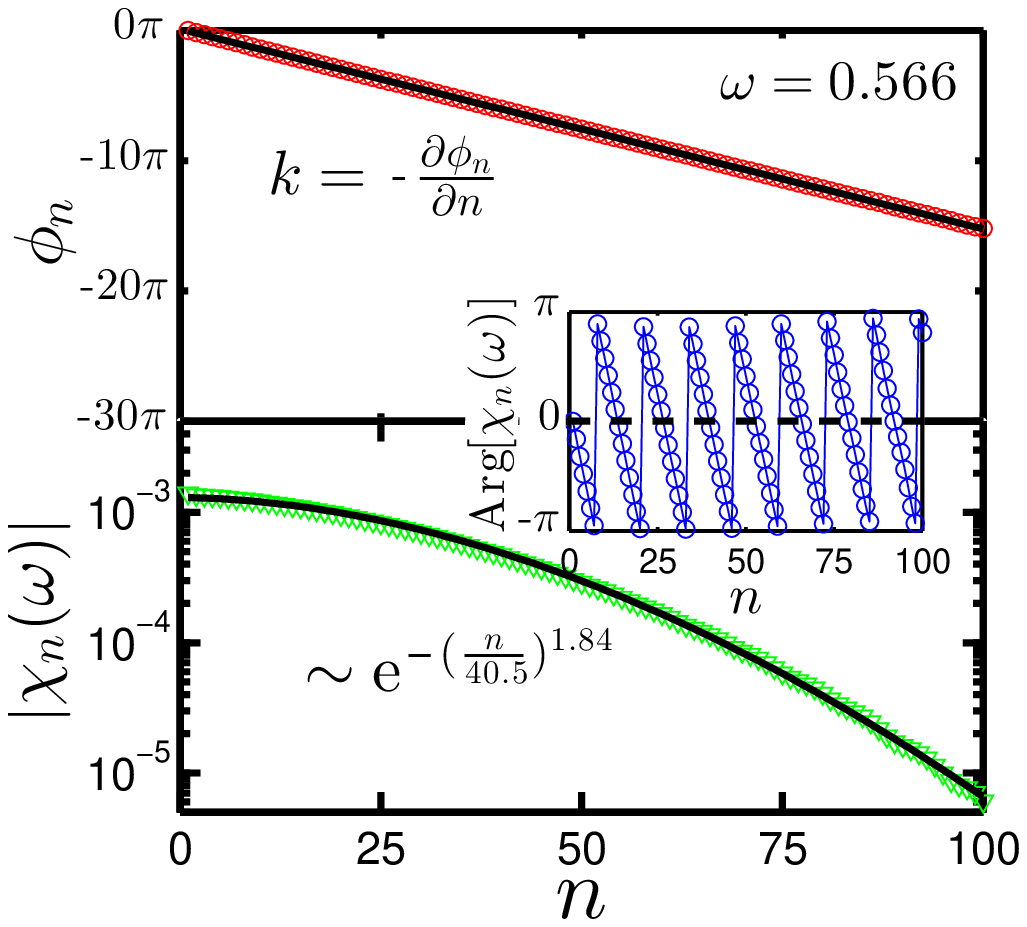}}
   \subfigure[\ The phase $\phi_n$]{
  \includegraphics[width=0.32\textwidth]{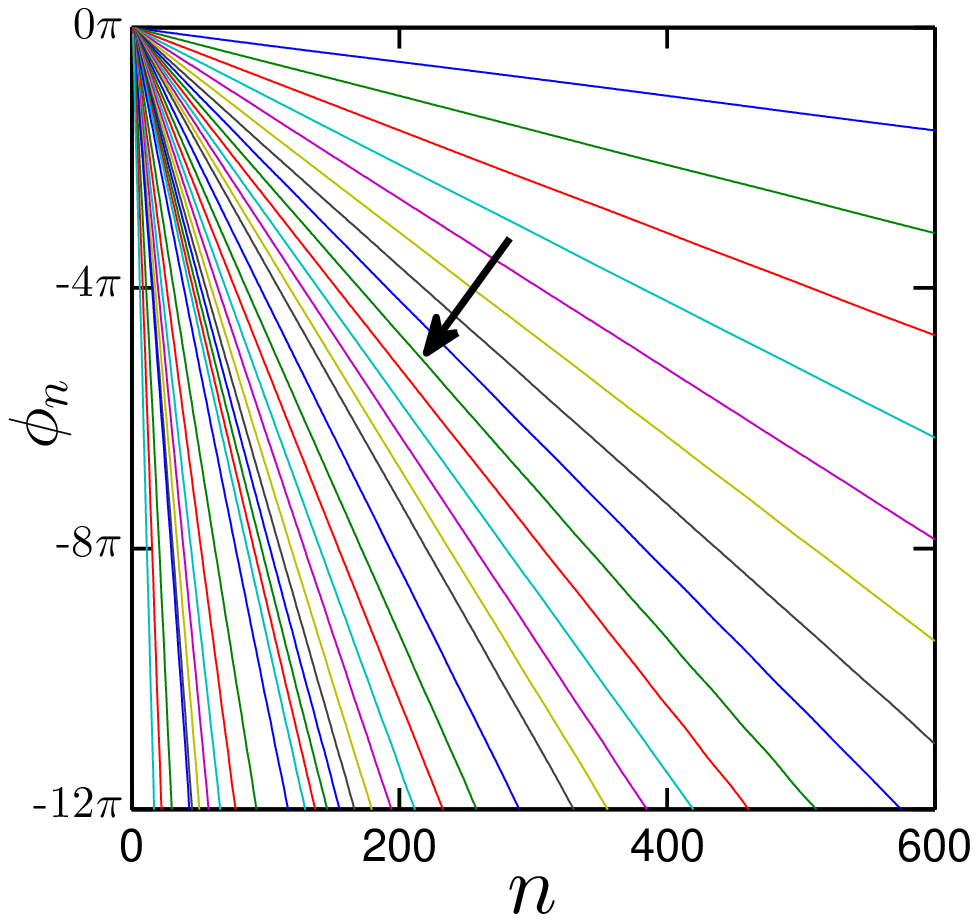}}
     \subfigure[\ the amplitude $|\chi_n|$]{
  \includegraphics[width=0.32\textwidth]{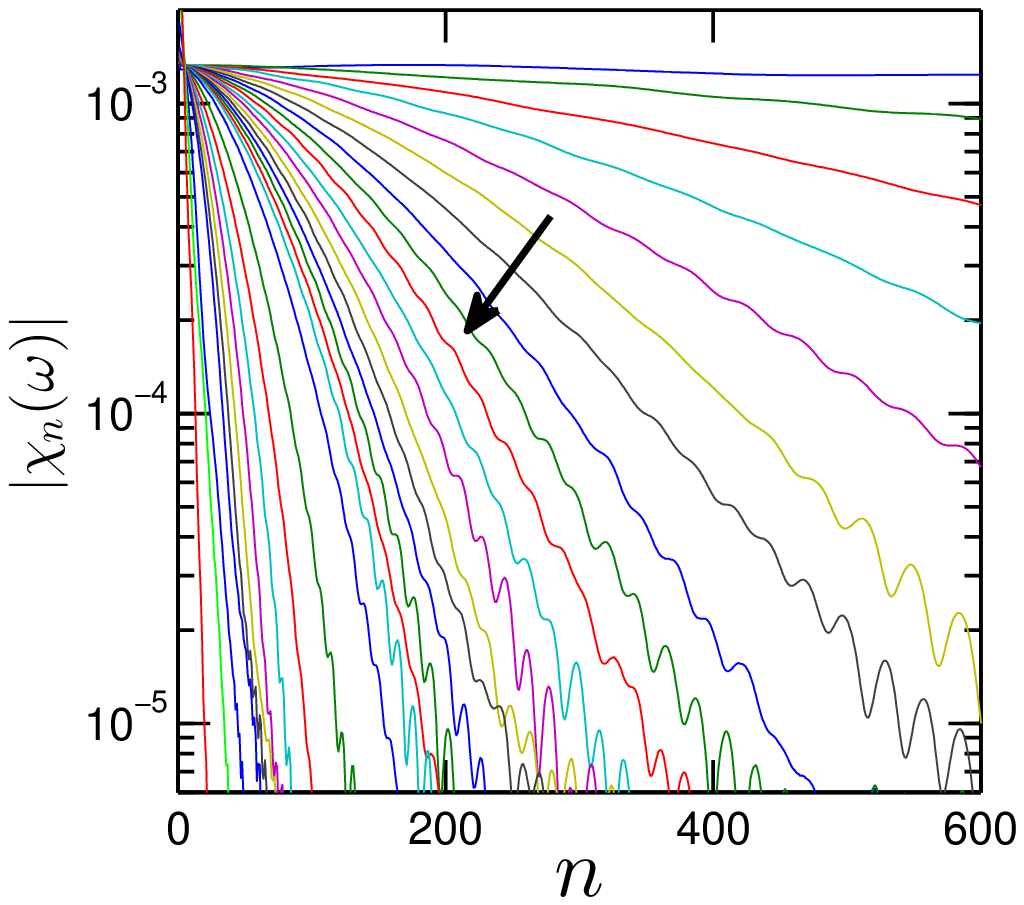}}
\caption{(color online) The response function for an
FPU-$\alpha\beta$ chain with length $N=2048$ at temperature $T=0.2$. (a) A detailed example for $\chi_n(\omega)$ at frequency $\omega= 0.566$. The upper panel shows the phase $\phi_n$ and the lower panel shows the amplitude $|\chi_n|$. The inset shows the principal values of the arguments
$\mathrm{Arg}[\chi_n]\in[-\pi,\pi)$. The principal
value jumps discontinuously by $2\pi$ when $-\pi$ is reached. To obtain a continuous varying phase
$\phi_n$, as depicted in the upper panel, we shift the arguments by 2$\pi$ after each such jump.
(b), (c) A comprehensive view of the phase and the amplitude for different frequencies, respectively, $\omega\in (0.0096, 2.387)$.
The driving frequency $\Omega$ increases along the arrow.
}
\label{fig:FPUab_all}
\end{figure*}

The \fpuab{} lattices containing a non-vanishing cubic term $V(\Delta x)= \frac{1}{2}\Delta
x^2+ \frac{1}{3} \Delta x^3 + \frac{1}{4}\Delta x^4$ and  $U(x)=0$  distinctly differ from \fpub{}
lattices. The inherent asymmetry of the interaction potential  yields a nonvanishing
internal pressure \cite{Zhong.12.PRE,Wang.13.PRE}.
Figure \ref{fig:FPUab_all} depicts the response function for the \fpuab{}
lattice.

As shown in \figref{fig:FPUab_all}(b), the perfect linear dependence of the phases
$\phi_n$ on $n$ can still be observed. Therefore, it corroborates the existence
of \AP{}'s in the \fpuab{} case.

\begin{figure}[b]
\includegraphics[width=0.9\columnwidth]{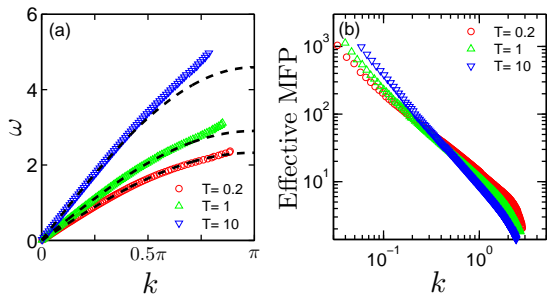}
\caption{(color online) (a) The dispersion relation for the \fpuab{} model at
different temperatures. The dashed curves are obtained from \eqref{eq:EPT} with $\alpha(T)$ set at
the sound speed as theoretically derived in Ref. \cite{Spohn.14.JSP}. (b) The effective MFPs.
}
\label{fig:afpu_dispersion}
\end{figure}

The amplitudes $|\chi_n|$, as shown in \figref{fig:FPUab_all}(c), however, deviate  from an exponential decay but instead depict multiple scales. Therefore, a strict MFP cannot be defined. Interestingly, as depicted as an example
in \figref{fig:FPUab_all}(a), their behavior can be fitted with a stretched exponential
\begin{equation}
|\chi_n(\omega)|=|\chi_1(\omega)| \exp\left(-\frac{(n-1)^a}{l^a} \right)
\end{equation}
with a frequency dependent parameter $a$.
 Therefore,
an effective single scale $\mfp_{\mathrm{eff}}$ can still be defined for each frequency $\omega$ if we average over all scales,
i.e.,
 \begin{equation}
 \mfp_{\mathrm{eff}}(\Omega):= \sum_1^\infty \frac{ |\chi_n(\Omega)|}{ |\chi_1(\Omega)|}
 \approx \int_0^\infty \exp \left(-\frac{n^a}{l^a}\right)dn
 \approx \frac{l}{a}\Gamma\left(\frac{1}{a}\right),
 \end{equation}
 where $\Gamma(x)$ is the Gamma function.
The dispersion relation and the so obtained effective MFPs are displayed in \figref{fig:afpu_dispersion}.

Although a renormalized phonon theory for \fpuab{} lattices does not exist,
we still find that Eq. \eqref{eq:EPT} holds approximately true for the dispersion of our \AP{}'s, see
the dashed lines in \figref{fig:afpu_dispersion}.
The corresponding sound speed $\alpha(T)$
matches well a recent result in \cite{Spohn.14.JSP}, which reads
\begin{equation}
\label{eq:spohn}
	\alpha^2= \frac{\frac{1}{2}\beta_T^{-2}+\mean{V+px;V+px}}
	{\beta_T(\mean{x;x}\mean{V;V}-
	\mean{x;V}^2)+\frac{1}{2}\beta_T^{-1}
\mean{x;x}},
\end{equation}
where $V(x)$ is the potential, $\mean{A;B}$ denotes the covariance
$\mean{AB}-\mean{A}\mean{B}$ for any two quantities $A$ and $B$, and
$p$ is the internal
pressure.


\subsection{  $\phi^4$-lattice. }
Here, the inter-particle potential is $V(\Delta x)=\hf \Delta x^2$
together with an on-site potential $U(x)=\frac{1}{4}x^4$.
For this momentum-nonconserving nonlinear lattice we still find that the phase follows a perfect linear decay. Moreover,
the corresponding MFP for the  \AP{} exists  with a single scale, i.e.,
$|\chi_n(\Omega)|$ nicely decays exponentially (figures for $\chi_n(\omega)$ are similar to Fig. (\ref{fig:R_all}) so they are not shown here).

\begin{figure}[t]
\includegraphics[width=0.9\columnwidth]{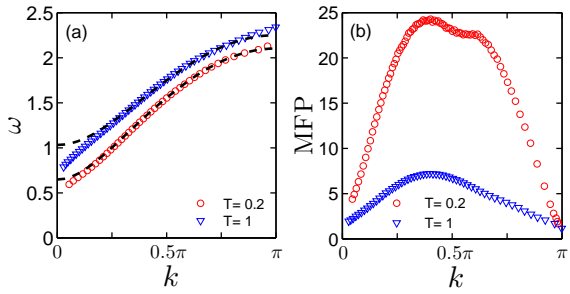}
\caption{(color online) (a) Phonon dispersion relation for a
$\phi^4$ nonlinear lattice at different temperatures. The dashed curves are obtained from
renormalized phonon
theory \cite{Li.13.PRE}. (b) Corresponding anharmonic phonon MFP-behavior.}
\label{fig:phi4}
\end{figure}

Our proposed \AP{} concept holds up  also in the presence of an onsite interaction.
The dispersion relation and the related MFPs are depicted  in \figref{fig:phi4}. Consistent with the validity of Fourier's law for momentum non-conserving systems the long
wavelength phonons exhibit finite MFPs \cite{Hu.00.PRE,Aoki.00.PLA}.  Our result for the
dispersion relation agrees well with the
renormalized phonon theory \cite{Li.13.PRE}, namely,
\begin{equation}
\omega=\sqrt{4\sin^2\frac{k}{2}+\sigma};\quad
\sigma=\frac{\sum_{i=1}^N \mean{x_i^4}}{\sum_{i=1}^N
 \mean{x_i^2}}.
 \end{equation}

\section{ Summary and discussion.}
The challenge of identifying phonon excitations in strongly nonlinear lattices beyond their
corresponding  harmonic approximation, termed here anharmonic phonons (\AP{}'s),  has been tackled with MD
via a theoretically
imposed ``tuning fork experiment''.
Doing so enables one to account for the role of
temperature, large frequencies and lattice nonlinearity in the strong nonlinearity regime.
This experiment-inspired phonon concept has been successfully tested over extended parameter regimes of frequency and temperature
for three archetype 1D nonlinear lattice models.
Note  that  a temperature dependent phonon MFP is typically not accessible with prior
theories without making
reference to additional assumptions
\cite{Callaway.59.PR,Holland.63.PR,Herring.54.PR,Pereverzev.03.PRE}.
Physically, the MFP relates to a phonon transport
relaxation time which  generally does {\it not} equal the phonon lifetime \cite{Sun.10.PRBa}.

Our concept for the MFP holds up beyond the validity regime of renormalized phonon theories
\cite{Alabiso.95.JSP,Alabiso.01.JPA,Gershgorin.05.PRL, Gershgorin.07.PRE,
Li.06.EL,Li.10.PRL}, depicting
a single, exponentially decaying scale for both
the \fpub{} lattice, exhibiting  anomalous heat
conduction, and the $\phi^4$ lattice, exhibiting  a Fourier's law behavior.  The case of the
\fpuab{} lattice turned out  intriguing in that the MFPs no longer exhibit a single scale but
decay with multiple scales.

A hallmark of our scheme is that the existence of the \AP{}'s and their MFPs is not postulated {\it
a priori}, but instead is physically corroborated by following the propagation of traveling waves with an
experiment. The outcome then either validates, or possibly also  invalidates, the existence of
\AP{} with a finite MFP. Our scheme thus distinctly differs from  existing phonon quasiparticle concepts  \cite{Henry.08.JCTN,Turney.09.PRB,Thomas.10.PRB,Sun.10.PRB,Zhang.14.PRL,Collins.78.PRB,Chaikin.book.2000}.
A most characteristic feature
within our scheme is that here we {\it directly}  search for the  existence of an \AP{} MFP.

The presented \AP{} concept allows one to characterize as well infinite
MFPs and multiple spatial decay scales, all being features that crucially impact anomalous thermal
transport  in low-dimensional  systems. This \AP{} concept
may as well spur interest in describing  heat transport in 3D materials, encompassing engineered complex materials, such as  phononic
metamaterials \cite{Yu.10.NN,Yang.14.NL,Davis.14.PRL}.

\begin{acknowledgments}
This work is supported by R-144-000-305-112 from MOE T2 (Singapore).
The authors would like to thank Dr. N. Li and Dr. J. Ren for useful discussions.
\end{acknowledgments}

\begin{thebibliography}{60}
\bibitem{Unit.00.NULL} Following the Appendix in Ref. \cite{Li.12.RMP}, we choose the atom mass $m$,
the lattice
constant $a$, the force constant $k_0$ and the Boltzmann constant $k_B$ as the four basic units to
scale all physical quantities involved to dimensionless quantities.
\end{thebibliography}

\end{document}